\documentclass[a4paper,12pt]{article}
\def\MARU#1{{\rm\ooalign{\hfil\lower.168ex\hbox{#1}\hfil\crcr\mathhexbox20D}}}
\newcounter{no}\newcommand{\NO}[1]{\setcounter{no}{#1}\Roman{no}}
\usepackage{amsmath}
\usepackage{bm}
\usepackage[dvips]{graphicx}

\begin{document}
\baselineskip = 6.5mm
\begin{center}
\begin{Large}
  {\bf Large $N$ behavior of string solutions \\ in the Heisenberg model}
\end{Large}

\vspace{1cm}
Takehisa Fujita,\footnote{fffujita@phys.cst.nihon-u.ac.jp}
Takuya Kobayashi\footnote{tkoba@phys.cst.nihon-u.ac.jp} and
Hidenori Takahashi\footnote{htaka@phys.ge.cst.nihon-u.ac.jp} $^*$

\vspace{5mm}
Department of Physics, Faculty of Science and Technology, \\ 
Nihon University, Tokyo, Japan

$ ^*$ Laboratory of Physics, Faculty of Science and Technology, \\ Nihon University, Chiba, Japan

\vspace{2cm}

{\bf ABSTRACT}

\vspace{10mm}
\begin{minipage}{15cm}
We investigate the large $N$ behavior of the complex solutions for the two magnon
system in the S=1/2 Heisenberg XXZ model. The Bethe ansatz equations 
are numerically solved for the string solutions with new iteration method. 
A clear evidence of the violation of the string configurations
is found at $N=22,62,121,200,299,417$, but the broken states are still Bethe states. 
The number of the Bethe states is consistent with the exact diagonalization, except 
one singular state.

\end{minipage}

\end{center}

\newpage
\section{Introduction}

In 1931, Bethe presented a unique and genius way of solving
one-dimensional Heisenberg model with spin $S=1/2$ systems \cite{bethe,katsura}.
The structure  of the Bethe ansatz solutions has been studied quite extensively.
Since the Bethe ansatz method has a history of many years,
the fundamental properties of the Bethe ansatz solution
for the Heisenberg XXZ model are well understood by now \cite{orb,yang2,tak,gau,fade_takh}.
However, the completeness of the Bethe ansatz solutions is still
discussed actively  \cite{koma_eza,kir_lis,fab_mcc}.
Bethe ansatz solutions contain complex solutions whose rapidities are complex
values.  For these solutions, Bethe proposed the {\it string hypothesis}.
He assumed the shape of the complex solutions which are called "string solutions"
 and showed that the real solutions together with such complex solutions give
 the correct number of the states \cite{bethe}.
Therefore, it is widely believed that the completeness of the Bethe's state holds true.

In two magnon system, the solutions are parametrized by two integer numbers $m_1$ and $m_2$.
The string solutions can be classified into two different types,
depending on the parity of $m_1+m_2$ in the two magnon state.
In fact, the behaviors of the wave function for the two string solutions are
quite different from each other. The string solutions with odd parity of
$m_1+m_2$ were first discussed by Bethe, and since then they have
been discussed quite extensively. In particular, Essler, Korepin and Schoutens
(EKS) carried out the careful study of the string solutions.
Here, we call this type of the string "EKS-string" as we define later in detail.
Recently, it is found that this EKS-string solutions break down to two real solutions
at $N=22,62,121 ...$  \cite{essl_kore_scho,isl_par,ila_kol_pal_pre},
which are partly predicted by Essler, Korepin and Schoutens.
On the other hand, there are other type of complex solutions which are treated
by  Vladimirov \cite{vlad}. This string solutions have  even parity of $m_1+m_2$,
but their  behaviors are quite different from the EKS-strings. 
Here, we call them "V-string". 
According to the string hypothesis, the imaginary part of the string is 1/2. 
However, Vladimirov shows that the imaginary part of the V-strings behaves like $\sqrt{N}$.

In this paper, we present the large $N$ behavior of the string solutions
of the Bethe ansatz equations for the two magnon system. Here, we have solved
the Bethe ansatz equations numerically up to quite large number of the site $N$.
The numerical evaluation has some difficulties for the complex solutions
in the Bethe ansatz equations. A simple iteration method cannot give any
convergent results, and thus we have to develope a new way to solve them.
This is what we have achieved here as we discuss it later in detail.
The solutions are compared with those of the exact diagonalization method up to $N=180$.
Here, we indeed confirm the violation of the EKS-strings up to $N=417$. 
Further, they are {\it not} out of the Bethe Ansatz solutions, and 
therefore, the number of the state is unchanged. 
In addition, we confirm that the  behavior of the V-string 
rapidity predicted by the $1/N$ expansion method \cite{vlad} is indeed consistent with our numerical solutions.

The paper is organized as follows. In the next section, we briefly explain 
the solutions of the Bethe ansatz equations. In section 3, we present 
a new iteration method of solving the Bethe equations numerically. In section 4, 
the string solutions for large $N$ cases are discussed and compared 
with the predictions of the $1/N$ expansion method. Section 5 summarizes what we have 
clarified.

\vspace{2cm}
%\newpage
\section{The solutions of the Bethe Ansatz equations }

The Heisenberg XXZ model with spin $1/2$ is a model which is most frequently studied 
among all the models of the spin systems. It is solved by the Bethe ansatz 
technique. Here, we briefly describe the Heisenberg model and the Bethe ansatz solutions. 
The XXZ model is described by the following Hamiltonian 
\begin{equation}
 H=J\sum_{i=1}^{N} \left( S_i^x S_{i+1}^x + S_i^y S_{i+1}^y + \Delta  S_i^z S_{i+1}^z  \right)
= \frac{J}{2} \sum_i \left(S_i^+ S_{i+1}^- + S_i^- S_{i+1}^+ + 2 \Delta S_i^z S_{i+1}^z\right),
\end{equation}
where $S_i^a$ is a spin operator at the site $i$ and $S_i^-(S_i^+)$ flips down (up) the spin,
and $\Delta$ is the anisotropic parameter. The periodicity $S_{N+i}=S_i$ is assumed.
For XXX model, we take $\Delta=1$.
This Hamiltonian can be diagonalized by the following Bethe state \cite{bethe,katsura} 
for two magnon system,
\begin{equation}
\left|\Psi_2 \right> = \sum_{x_1<x_2} A(x_1,x_2) S^-_{x_1} S^-_{x_2} \left|0\right>,
\end{equation}
where $\left|0\right>$ is the ferromagnetic state with all spins up.
 The coefficient $ A(x_1,x_2)$ is assumed to be of the following shape,
\begin{equation}
 A(x_1,x_2)= e^{ik_1 x_1+ik_2 x_2 } +  e^{ik_2 x_1+ik_1 x_2} e^{-i\varphi}
\end{equation}
and must satisfy
\begin{equation}
A(x_j,x_j)+ A(x_j+1,x_j+1)-2A(x_j,x_j+1)=0.
\end{equation}
Therefore, the phase shift $\varphi$ should satisfy the following equation,
\begin{equation}
\cot {\varphi\over 2}= \dfrac{\Delta \sin \dfrac{k_1-k_2}{2}}
{\cos \dfrac{k_1+k_2}{2}-\Delta\cos \dfrac{k_1-k_2}{2}},
\label{bethe_phi_eq}
\end{equation}
where $\varphi$ is taken to be $ -\pi \leq \varphi \leq \pi $. Further, imposing 
the periodic boundary conditions, we have
\begin{eqnarray}
k_1 N + \varphi &= 2\pi m_1,
\label{bethe_rap_eq1} \\
k_2 N - \varphi &= 2\pi m_2,
\label{bethe_rap_eq2}
\end{eqnarray}
where $m_1$ and $m_2$ are integers running between $0$ and $N-1$. 
Without loss of generality, we can take $m_1 \le m_2$. 
In this case, the energy eigenvalue of the Hamiltonian can be written as
\begin{equation}
 E/J= \sum_{j=1}^2 \cos k_j + \Delta \left({N \over 4}-2 \right).
\end{equation}
We can define the rapidity as
\begin{equation}
\lambda_j = - \dfrac{1}{2} \cot \dfrac{k_j}{2}.
\end{equation}
In this case, the energy is given by
\begin{equation}
E/J=-\dfrac{1}{2} \sum_{j=1}^2 \dfrac{1}{\lambda^2_j+1/4} + \Delta \left(\frac{N}{4} -2 \right)+2.
\end{equation}
In order to obtain the energy eigenvalues of the Hamiltonian,
one has to solve eqs.(\ref{bethe_phi_eq}), (\ref{bethe_rap_eq1}) and (\ref{bethe_rap_eq2}).
Here, we show  the rapidity and  the energy in the case of $N=8$ 
and $\Delta=1$ for simplicity. We must solve the equation
\begin{equation}
 \cos \dfrac{\pi}{N}(m_1+m_2) \cos \left( \dfrac{NK}{2}-{\pi\over 2}(m_1-m_2) \right)
   = \cos \left( \dfrac{N-2}{4}K-{\pi\over 2}(m_1-m_2)\right),
\end{equation}
where $K=(k_1-k_2)/2$.
This equation can be solved analytically once we specify the values of $m_1$ and $m_2$.
It is difficult to prove the completeness of the Bethe Ansatz solutions.
Therefore, we must carefully treat the Bethe Ansatz equation whose solutions are indeed
the answer of the model.

In table \ref{tbl_n8d1}, we show the calculated results of the energies
for $N=8$ with the two down spins together with the energies by the exact diagonalization
for $\Delta=1$. The detailed method to solve the Bethe equations numerically 
will be given in section 3. 

As can be seen from  table \ref{tbl_n8d1}, there is one state, 
denoted by $\bm{*}$, which cannot be reproduced by the Bethe ansatz. 
The configurations of the states are different for different values of  $\Delta$. 
Here, we only show the results for $\Delta=1$ case. For other values of $\Delta$, we 
only make a comment here.  We have the real solution for  $\Delta =1$ in the Category I, 
which is defined in \cite{orb}. 
The string solutions appear for the Category II and III. 
For $\Delta=2$, there is an irregular solution at $(m_1, m_2)=(0,7)$. 
Even though the element belongs to the Category I, the solution is a string. 
We have a similar configuration for the case $\Delta > 1$. 
On the other hand, there is no complex solution for $\Delta=1/2$ at $N=8$. 
In what follows, we will only discuss for  $\Delta=1$ case. The general 
 $\Delta$ cases will be treated elsewhere. 

Now, there is a state which cannot be reproduced by the Bethe ansatz solution 
for $\Delta =1$. We can easily verify that the energy of this state is given by
\begin{equation}
E/J = \Delta \left(\dfrac{N}{4}-1 \right).
\end{equation}
The corresponding state is
\begin{equation}
\left|Non-Bethe \right> = \sum_{i=1}^N (-)^i S_i^- S_{i+1}^- \left|0\right>.
\label{wavefunc_nonbethe}
\end{equation}
Note that the {\it kinetic term} of this state is null,
\begin{equation}
\sum_i \left(S_i^+ S_{i+1}^- + S_i^- S_{i+1}^+ \right) \left|Non-Bethe \right> = 0.
\end{equation}
Thus, we cannot describe this state in the Bethe Ansatz equation because 
eq.(\ref{bethe_phi_eq}) is not valid for this case any more. 
We  note here that this state is not reproduced by the algebraic Bethe Ansantz 
at the isotropic point either \cite{sidd}.

\begin{table}
\begin{center}
\caption{
The energies $E_{exact}$ and $E_{Bethe}$ which are solved by the exact diagonalization
and by the Bethe ansatz method are presented for the  $N=8$ and  $\Delta=1$ case.
$*$ means that there is no corresponding state in the Bethe Ansatz method.
}
\label{tbl_n8d1}

\vspace{2mm}
{\scriptsize
\begin{tabular}{cccccc}
\hline
$E_{exact}/J$ & $E_{Bethe}/J$ & ($m_1,m_2$)  & $k_1$ & $k_2$ & $\varphi$  \rule[-3mm]{0pt}{8mm} \\
\hline
\hline
$-1.801938$ & $-2\cos \dfrac{\pi}{7}$ & (3,5) & $6\pi/7$ & $8\pi/7$ & $6\pi/7$
     \rule[-3mm]{0pt}{8mm} \\
$-1.267035$ & $-1.267035$ & (2,4) & $0.6035\pi$ & $0.8965\pi$ & $0.8279\pi$  \rule[0mm]{0pt}{7mm} \\
& & (4,6) & $1.1035\pi$ & $1.3965\pi$ & $0.8279\pi$ \\
$-1.144123$ & $-\dfrac{\sqrt{10}+\sqrt{2}}{4}$ & (2,5) & $0.5875\pi$ & $1.1625\pi$ & $0.7003\pi$
     \rule[0mm]{0pt}{7mm} \\
& & (3,6) & $0.8375\pi$ & $1.4125\pi$ & $0.7003\pi$ \\
$-0.4450419$ & $-2\cos \dfrac{3}{7}\pi$ & (2,6) & $4\pi/7$ & $10\pi/7$ & $4\pi/7$ \rule[0mm]{0pt}{7mm} \\
$-0.4370160$ & $-\dfrac{\sqrt{10}-\sqrt{2}}{4}$ & (1,4) & $0.3184\pi$ & $0.9316\pi$ & $0.5475\pi$
     \rule[0mm]{0pt}{7mm} \\
& & (4,7) & $1.0684\pi$ & $1.6816\pi$ & $0.5475\pi$ \\
$-0.2586520$ & $-0.2586520$ & (1,5) & $0.3085\pi$ & $1.1915\pi$ & $0.4684\pi$  \rule[0mm]{0pt}{7mm} \\
& & (3,7) & $0.8085\pi$ & $1.6915\pi$ & $0.4684\pi$ \\
$0$          & 0 & (0,4) & 0 & $\pi$ & $0$  \rule[0mm]{0pt}{7mm} \\
& & (1,3) & $\pi/3$ & $2\pi/3$ & $2\pi/3$ \\
& & (5,7) & $4\pi/3$ & $5\pi/3$ & $2\pi/3$ \\
 0.2928932   & $\dfrac{2-\sqrt{2}}{2}$ & (0,3) & 0 & $3\pi/4$ & 0  \rule[0mm]{0pt}{7mm} \\
& & (0,5) & 0 & $5\pi/4$ & 0 \\
 0.4370160   & $\dfrac{\sqrt{10}-\sqrt{2}}{4}$ & (1,6) & $0.2990\pi$ & $1.4510\pi$ & $0.3920\pi$
     \rule[0mm]{0pt}{7mm} \\
& & (2,7) & $0.5490\pi$ & $1.7010\pi$ & $0.3920\pi$ \\
 1.0    & 1 & (0,2) & 0 & $\pi/2$ & 0 \rule[0mm]{0pt}{7mm} \\
 & & (0,6) & 0 & $3\pi/2$ & 0 \\
 $\bm{1.0}$    & $\bm{*}$ & $\bm{*}$ & & &   \rule[0mm]{0pt}{7mm} \\
 $1.144123$ & $\dfrac{\sqrt{10}+\sqrt{2}}{4}$ & (1,2) & $3\pi/8 + 0.9578i$ &
    $3\pi/8- 0.9578i$ & $\pi+ 7.6626i$   \rule[0mm]{0pt}{7mm} \\
 & & (6,7) & $13\pi/8 + 0.9578i$ & $13\pi/8 - 0.9578i$ & $\pi + 7.6626i$ \\
 1.246980   & $-2\cos \dfrac{5}{7}\pi$ & (1,7) & $2\pi/7$ & $12\pi/7$ & $2\pi/7$ \rule[0mm]{0pt}{7mm} \\
 1.525687   & 1.525687 & (1,1) & $\pi/4+ 0.3945i$ & $\pi/4- 0.3945i$ & $ 3.1559i$ \rule[0mm]{0pt}{7mm} \\
 & & (7,7) & $7\pi/4+ 0.3945i$ & $7\pi/4- 0.3945i$ & $ 3.1559i$ \\
 1.707107   & $\dfrac{2+\sqrt{2}}{2}$ & (0,1) & 0 & $\pi/4$ & 0  \rule[0mm]{0pt}{7mm} \\
 & & (0,7) & 0 & $7\pi/4$ & 0 \\
 2.0   & 2 & (0,0) & 0 & 0 & 0  \rule[-3mm]{0pt}{9mm} \\
 \hline
 \rule{15mm}{0pt} & \rule{20mm}{0pt} & \rule{20mm}{0pt} & \rule{15mm}{0pt}& \rule{15mm}{0pt}&\rule{15mm}{0pt} \\
\end{tabular}
}
\end{center}
\end{table}

Bethe suggested that there is a string solution whose imaginary part is 
infinity \cite{bethe}. 
However, the wave function is divergent. 
For appropriate normalization, we get the wave function 
given as eq.(\ref{wavefunc_nonbethe}). 
Therefore, the wave function eq.(\ref{wavefunc_nonbethe}) corresponds to the singular Bethe state.

We comment on the degeneracy of the energy states.
Suppose that the $k_1, k_2$ are solutions for the configuration $(m_1,m_2)$ ($1 \le m_1 \le m_2 \le N-1$)
 with the phase $\varphi$. The Bethe equations are
\begin{equation}
k_1 N + \varphi = 2 \pi m_1, \quad k_2 N - \varphi = 2 \pi m_2.
\end{equation}
Now we consider the configuration $(N-m_2, N-m_1)$ with $k_1^\prime, k_2^\prime, \varphi^\prime$.
We can easily show that two states have the same energy if the following equations are 
satisfied 
\begin{equation}
k_1^\prime + k_2^\prime = 4\pi - (k_1+k_2), \quad k_1^\prime - k_2^\prime = k_1-k_2,
\quad \varphi^\prime= \varphi.
\end{equation}
Note that
\begin{equation}
k_i^\prime = k_i \;(i=1,2), \quad \textrm{for } \quad m_1+m_2=N.
\end{equation}
In this case, there is no pair and the Bethe state is a single state.
For $(0,m_2), (1 \le m_2 \le N-1)$ case, the configuration $(0,N-m_2)$ is a solution 
with the same energy as the configuration $(0,m_2)$ when
\begin{equation}
k_1^\prime + k_2^\prime = 2\pi - (k_1+k_2), \quad k_1^\prime - k_2^\prime = -(k_1-k_2)-2\pi,
\quad \varphi^\prime+\varphi=0.
\end{equation}

\vspace{2cm}
%\newpage
\section{New iteration method }

The Bethe ansatz equations (5),(6) and (7) can be solved by the iteration 
method even for the large $N$ cases. The convergence of the iteration method 
is good for the normal cases, but the complex solutions cannot be obtained 
easily by the simple-minded iteration method. 
Here, we present somewhat a different way of numerically solving 
the Bethe equations. But before going to the new method, we briefly 
explain the usual iteration method. %Now,
The $\varphi$ in eq.(5) can be written as
\begin{equation}
 \varphi=2\cot^{-1}\frac{\Delta\sin\dfrac{k_1-k_2}2}
{\cos\dfrac{k_1+k_2}2-\Delta\cos\dfrac{k_1-k_2}2}
\equiv X(\varphi).
\end{equation}
The usual iteration method is done as follows
\begin{equation}
\varphi^{(i)}=X(\varphi^{(i-1)}),
\end{equation}
where we start from some initial value of $\varphi^{(0)} $. 
$k_1$ and $k_2$ are given in eqs.(6) and (7). As mentioned above, this 
procedure works well for the normal cases. But for the complex solutions, 
it does not work. 

In this paper, we show a new way of solving the complex solutions. 
First, we define a new variable $v$ by 
\begin{align}
 \varphi &= iv \quad\quad\quad {\rm for} \ \ \ (m,m) \quad\quad\quad \textrm{(V-string)}, \\
 \varphi &= \pi+iv \quad {\rm for} \ \ \  (m,m+1)  \quad \textrm{(EKS-string)},
\end{align}
where $v$ is taken to be  real. 
For V-string case $(m,m)$, we obtain 
\begin{equation*}\begin{split}
\coth\frac{v}2 &=\frac{-\Delta\sinh\dfrac{v}{N}}{\cos\dfrac{2m\pi}N
-\Delta\cosh\dfrac{v}N}, \\
\sinh\frac{v}2 &=\frac{\cos\dfrac{2m\pi}N-\Delta\cosh\dfrac{v}N}
{-\Delta\sinh\dfrac{v}{N}}\cosh\dfrac{v}2  \equiv X(v), \\
v &=2 \ln\left(X+\sqrt{X^2+1}\right).
\end{split}\end{equation*}
Therefore, the iteration equation becomes
\begin{equation}
v^{(i)}=2\ln\left(X+\sqrt{X^2+1}\right)  \Big|_{v=v^{(i-1)}}.
\end{equation}
On the other hand, for the case of EKS-string $(m,m+1)$, we have 
\begin{equation*}\begin{split}
\tanh\frac{v}2 &=\frac{-\Delta\sinh\dfrac{v}{N}}{\cos\dfrac{(2m+1)\pi}N
-\Delta\cosh\dfrac{v}N}, \\
\cosh\frac{v}2 &=\frac{\cos\dfrac{(2m+1)\pi}N-\Delta\cosh\dfrac{v}N}
{-\Delta\sinh\dfrac{v}{N}}\sinh\dfrac{v}2  \equiv Y(v), \\
v &=2 \ln\left(Y+\sqrt{Y^2-1}\right).
\end{split}\end{equation*}
Therefore, the iteration equation becomes 
\begin{equation} v^{(i)}=2\ln\left(Y+\sqrt{Y^2-1}\right) \Big|_{v=v^{(i-1)}}.
\end{equation}
We can carry out the iteration procedure for million times, and 
obtain good convergent results. 
In this way, this iteration method gives complete solutions, and thus
we can find all of the string solutions of V-string as well as 
EKS-string. With this method, we can find the string solutions 
up to $N \simeq 5000$.

\vspace{2cm}

\section{String solutions for large $N$}

\begin{figure}
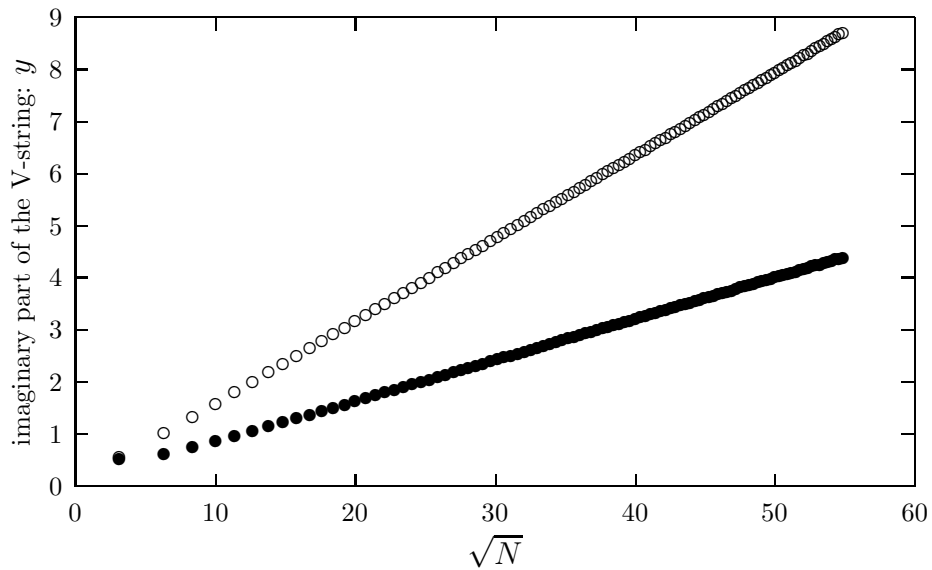

\begin{center}
\caption{
The large $N$ behavior of the V-string solutions for the XXX model ($\Delta=1$). Plot of the imaginary part
of the string against $\sqrt{N}$ for $(m_1,m_2)=(1,1)$ (plotted with the symbol $\circ$) and
 $(m_1,m_2)=(2,2)$ (plotted with the symbol $\bullet$).
}
\label{fig_im_string_m1}
\setlength{\unitlength}{0.240900pt}
\ifx\plotpoint\undefined\newsavebox{\plotpoint}\fi
\sbox{\plotpoint}{\rule[-0.200pt]{0.400pt}{0.400pt}}%
%
\end{center}
\end{figure}

Next, we discuss the properties of the string solutions. 
For the string hypothesis,  the complex solutions are given  
for $M$-magnon system by
\begin{equation}
\lambda_j = x + i \left( \dfrac{M+1}{2} - j \right) + O(e^{-\delta N}), \quad \quad (j=1,2, \cdots M),
\end{equation}
where $x$ is the same real part of the complex roots $\lambda_j$, 
and $\delta$ is a positive parameter.
According to the string hypothesis, the imaginary part $y$ of the string 
at large $N$ is given as, 
\begin{equation}
y \rightarrow \dfrac{1}{2}
\end{equation}
for the two magnon case ($M=2$). On the other hand, 
Vladimirov predicts \cite{vlad}
the non-string type complex solutions by making use of the $1/N$ 
expansion method. Defining the real and imaginary parts of the rapidity 
by $\lambda=x \pm i y$, they can be expressed as
\begin{equation}
 x=\dfrac{N}{\pi \ell} \left(1 - \dfrac{1}{N} - \dfrac{\pi^2 \ell^2}{6 N^2} + \cdots \right), \quad \quad
 y=\dfrac{\sqrt{N}}{\pi \ell} \left[1+ \dfrac{1}{N} \left(\dfrac{\pi^2 \ell^2}{24} - \dfrac{1}{2}\right)
  + \cdots \right],
\end{equation}
where $\ell = 2,4,6, \cdots \le \sqrt{N}$. For large $N$, they behave as
\begin{equation}
x \sim N, \quad \quad y \sim \sqrt{N}.
\label{large_n_vlad}
\end{equation}
In table \ref{tbl_string}, we show the complex solutions of the XXX model 
for several cases of  $N$. 
The Vladimirov's solutions are given in the following  configurations,
\begin{equation}
 (m_1, m_2) = (r, r), \quad\quad r = 1,2, \cdots \left[\dfrac{\sqrt{N}}{2}\right],
\end{equation}
where $[x]$ denotes the maximum integer $n$  for $n \le x$ (Gauss's symbol).
In figures \ref{fig_im_string_m1} and \ref{fig_re_string_m1},
 we show the large $N$ behavior of the imaginary part and the real part 
of the complex solutions, respectively.
We can see that the large $N$ behavior of the configuration for $(m_1,m_2)=(1,1),(2,2)$ is indeed consistent with eq. (\ref{large_n_vlad}).

\begin{table}
\begin{center}
\caption{
The energies $E_{exact}$ and the string solutions and non-string type solutions which are solved numerically
 for the XXX model ($\Delta=1$). $\lambda_{vlad}$ is the Vladimirov's prediction.
}
\label{tbl_string}

\vspace{2mm}
{\scriptsize
\begin{tabular}{ccccl}
\hline
$N$ & $E_{exact}/J$ & ($m_1,m_2$) & $\lambda$ & \multicolumn{1}{c}{$\lambda_{vlad}$}  \rule[-3mm]{0pt}{8mm} \\
\hline \hline
12 & 2.76225903 & (1,1) & $1.6510879 \pm 0.619417i$ & $1.6538681 \pm 0.617998 i\;\; (\ell=2)$ \rule[0mm]{0pt}{6mm}\\
   & 2.25054560 & (2,2) & $0.5769306 \pm 0.500242i$ & \multicolumn{1}{c}{$-$} \\
   & $\vdots$  & $\vdots$ & $\vdots$ &  \\
   & 2.4909371 & (1,2) & $1.0184604 \pm 0.4808314i$ & \multicolumn{1}{c}{$-$} \\
   & 2.0669861 & (2,3) & $0.2679495 \pm 0.4999999i$ & \multicolumn{1}{c}{$-$} \\
   & $\vdots$ & $\vdots$ & $\vdots$ & \rule[-3mm]{0pt}{6mm}  \\
\hline
22 & 5.4233830 & (1,1) & $3.2913953 \pm 0.7913153i$ & $3.2918067 \pm 0.7910191i\;\; (\ell=2)$  \rule[0mm]{0pt}{6mm}\\
   & 5.2100924 & (2,2) & $1.5434373 \pm 0.5190729i$ & $1.5531493 \pm 0.5155661i\;\; (\ell=4)$ \\
   & 4.9289111 & (3,3) & $0.8664011 \pm 0.500090 i$ & \multicolumn{1}{c}{$-$} \\
   & $\vdots$  & $\vdots$ & $\vdots$ &  \\
   & 5.0704584 & (2,3) & $1.1559456 \pm 0.4978173i$ & \multicolumn{1}{c}{$-$} \\
   & 4.7922902 & (3,4) & $0.6426630 \pm 0.4999987i$ & \multicolumn{1}{c}{$-$} \\
   & 4.5793732 & (4,5) & $0.2936265 \pm 0.5000000i$ & \multicolumn{1}{c}{$-$} \\
   & $\vdots$ & $\vdots$ & $\vdots$ & \rule[-3mm]{0pt}{6mm}  \\
\hline
60 & 14.989239 & (1,1) & $9.3722865 \pm 1.2576148i$ & $9.3723056 \pm 1.2575917i\;\; (\ell=2)$ \rule[0mm]{0pt}{6mm}\\
   & 14.957218 & (2,2) & $4.6566352 \pm 0.6881097i$ & $4.6571019 \pm 0.687560i \;\; (\ell=4)$ \\
   & 14.904929 & (3,3) & $3.0641980 \pm 0.5397430i$ & $3.0673519 \pm 0.5372569i\;\; (\ell=6)$ \\
   & 14.834704 & (4,4) & $2.2441546 \pm 0.5042061i$ & \multicolumn{1}{c}{$-$} \\
   & $\vdots$ & $\vdots$ & $\vdots$ &  \\
   & 14.931927 & (2,3) & $3.7935172 \pm 0.1297153i$ & \multicolumn{1}{c}{$-$} \\
   & 14.871226 & (3,4) & $2.6121248 \pm 0.4812710i$ & \multicolumn{1}{c}{$-$} \\
   & 14.793839 & (4,5) & $1.9631184 \pm 0.4990019i$ & \multicolumn{1}{c}{$-$} \\
   & $\vdots$ & $\vdots$ & $\vdots$ & \rule[-3mm]{0pt}{6mm}  \\
\hline
100 & 24.996095 & (1,1) & $15.745726 \pm 1.6103627i$ & $15.745730 \pm 1.6103563i\;\; (\ell=2)$ \rule[0mm]{0pt}{6mm}\\
    & 24.984411 & (2,2) & $7.8560250 \pm 0.8483587i$ & $7.8561233 \pm 0.8481972i\;\; (\ell=4)$ \\
    & 24.965063 & (3,3) & $5.2162758 \pm 0.6205981i$ & $5.2169764 \pm 0.6195799i\;\; (\ell=6)$ \\
    & 24.938297 & (4,4) & $3.8857124 \pm 0.5339379i$ & $3.8883774 \pm 0.5314536i\;\; (\ell=8)$ \\
    & 24.904572 & (5,5) & $3.0756442 \pm 0.5062334i$ & $3.0816823 \pm 0.5074819i\;\; (\ell=10)$ \\
    & 24.864499 & (6,6) & $2.5254416 \pm 0.5006816i$ & \multicolumn{1}{c}{$-$} \\
    & $\vdots$ & $\vdots$ & $\vdots$ &  \\
    & 24.952091 & (3,4) & $4.5042779 \pm 0.3353532i$ & \multicolumn{1}{c}{$-$} \\
    & 24.922034 & (4,5) & $3.4477639 \pm 0.4797970i$ & \multicolumn{1}{c}{$-$} \\
    & 24.885222 & (5,6) & $2.7784399 \pm 0.4976799i$ & \multicolumn{1}{c}{$-$} \\
    & $\vdots$ & $\vdots$ & $\vdots$ &   \rule[-3mm]{0pt}{6mm}\\
\hline
\end{tabular}
}
\end{center}
\end{table}

\begin{table}
\begin{center}
\caption{
The energies and the string solutions near the breaking point for the XXX model.
$*$ means the exact diagonalization is not possible. }
\label{tbl_string_break}

\begin{tabular}{ccccc}
\hline
$(m_1,m_2)$ & $N$ & $E_{exact}/J$ & $E_{bethe}/J$  & $\lambda$  \rule[-2mm]{0pt}{7mm}\\
\hline\hline
$(1,2)$ & 21 & $5.05193773$ & $5.05193773$ & $2.1766735 \pm 0.17888955i$  \rule{0pt}{5mm}\\
        & 22 & $5.31910954$ & $5.31910954$ & 2.3727865 \\
        &    &  &              & 2.2284619 \\
\hline
$(2,3)$ & 61 & $15.1840740$ & $15.1840740$ & $3.8607743 \pm 0.06641234i$  \rule{0pt}{5mm}\\
        & 62 & $15.4361213$ & $15.4361213$ & 4.0193617 \\
        &    &  &              & 3.8366348 \\
\hline
$(3,4)$ &120 & $29.9665138$ & $29.9665138$ & $5.4408068 \pm 0.05989996i$  \rule{0pt}{5mm}\\
        &121 & $30.2170566$ & $30.2170566$ & 5.5399897 \\
        &    &  &              & 5.4351240 \\
\hline
$(4,5)$ &199 & $*$ & $49.7298474$ & $7.0260588 \pm 0.04464858i$  \rule{0pt}{5mm}\\
        &200 & $*$ & $49.9800466$ & 7.1045560 \\
        &    &   &              & 7.019490  \\
\hline
\end{tabular}
\end{center}
\end{table}

\begin{figure}
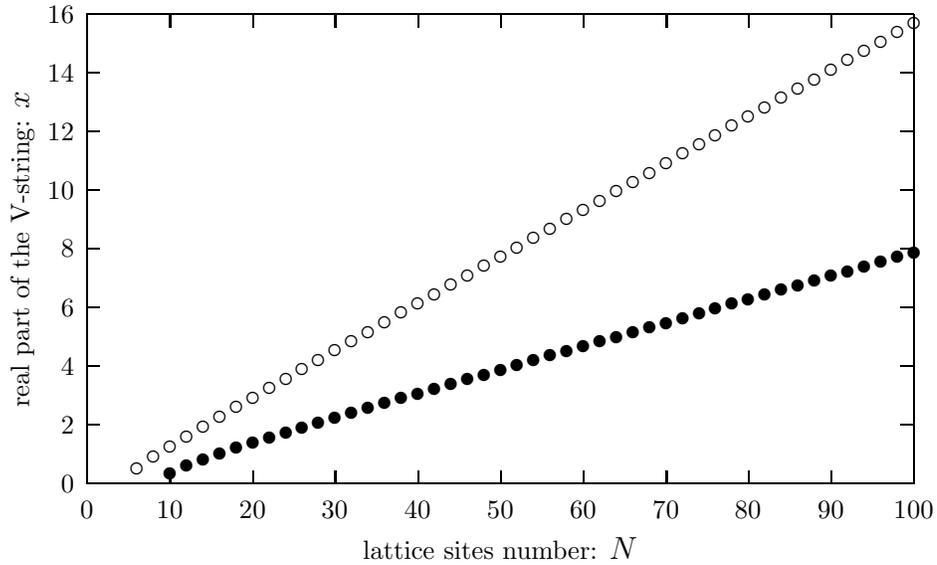

\begin{center}
\caption{
The large $N$ behavior of the EKS-string solutions for the XXX model ($\Delta=1$). Plot of the imaginary part
of the string against the sites number $N$ for $m_2=m_1+1$ case.
}
\label{fig_im_string_m11}
\begin{minipage}{12.5cm}
\setlength{\unitlength}{0.240900pt}
\ifx\plotpoint\undefined\newsavebox{\plotpoint}\fi
\sbox{\plotpoint}{\rule[-0.200pt]{0.400pt}{0.400pt}}%
%
\end{minipage}
\end{center}
\end{figure}

On the other hand, the large $N$ behavior of the configurations for $m_2=m_1+1$ case is different.
We show the behavior of the imaginary part and the real part of the complex solutions in figure
\ref{fig_im_string_m11} and figure \ref{fig_re_string_m11}.
There are branch points where the string solutions break down.
  At the branch point, the imaginary part is close to zero, and  the real part is divided
into two parts. However, the broken string becomes two real solutions 
in the Bethe Ansatz equation and thus 
 these broken states are still degenerate, which is shown in table \ref{tbl_string_break}. 
In table \ref{tbl_conf_string}, we list the configurations 
for the complex solutions. 

\begin{figure}
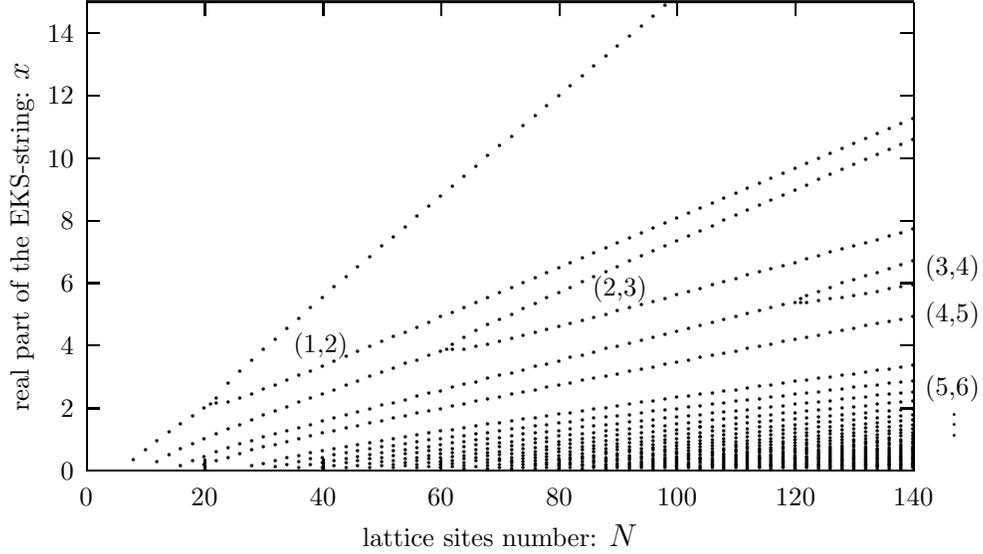

\begin{center}
\caption{
 Plot the minimum($s_0$) of $m_1$  against  the sites number $N$ for $m_2=m_1+1$ case in the XXX model.
 $\circ$ means that the point is excluded.
The fitting function which is denoted by dotted line is given by $s_0 \sim N^{0.4934}$.
}
\label{fig_minimum_m1}
\begin{minipage}{14.5cm}
\setlength{\unitlength}{0.240900pt}
\ifx\plotpoint\undefined\newsavebox{\plotpoint}\fi
\sbox{\plotpoint}{\rule[-0.200pt]{0.400pt}{0.400pt}}%
%
\end{minipage}
\end{center}
\end{figure}

In figure \ref{fig_minimum_m1}, the minimum number of $m_1$ and 
the sites number $N$ are given. 
 The number of the string does not increase linealy for $N$.
Note that, in both cases, there are string solutions whose imaginary part
is $1/2$. However, the value of the imaginary part starts to deviate from 
$1/2$ when the value of the site number $N$ increases, and finally 
becomes zero for sufficiently large values of $N$.

Finally, the string configurations for the XXX model ($\Delta=1$) are classified by the
large $N$ behavior. In \cite{ila_kol_pal_pre}, Ilakovac et al. 
distinguish the string solution in terms of
the parity of $m_1+m_2$. They call the string solutions as $s$-string 
for odd parity case, and $c$-string for even parity case. 
The string solutions are given by
\begin{equation}
\textrm{V-string($c$-string):} \quad (m_1,m_2)=(r, r), \; (N-r,N-r), \quad \quad \left(0<r<\dfrac{N}{4}\right),
\end{equation}
\begin{equation}
\textrm{EKS-string($s$-string):} \quad (m_1,m_2)=(s, s+1), \; (N-s-1,N-s), \quad \quad
   \left(s_0 < s < \dfrac{N}{4} \right),
\end{equation}
where $r, s$ are intger and $s_0$ is minimum of $m_1$. 
The distribution of $s_0$ is given by figure \ref{fig_minimum_m1}.
For the large $N$, we have
\begin{equation}
 s_0 \sim 0.3347 N^{0.4934} \sim \sqrt{N}.
\end{equation}
Therefore, we can estimate the ratio of the number of the string  $N_{\textrm{string}}$ and
the total states $N_{\textrm{total}}=N(N-1)/2$ at
\begin{equation}
\dfrac{N_{\textrm{string}}}{N_{\textrm{total}}} \sim \dfrac{2}{N}-\dfrac{2}{N \sqrt{N}}.
\label{ratio_pre}
\end{equation}
If the string violation does not occur, the ratio is
\begin{equation}
\dfrac{N_{\textrm{string}}}{N_{\textrm{total}}} \sim \dfrac{2}{N}-\dfrac{6}{N^2}.
\label{ratio_sthyp}
\end{equation}

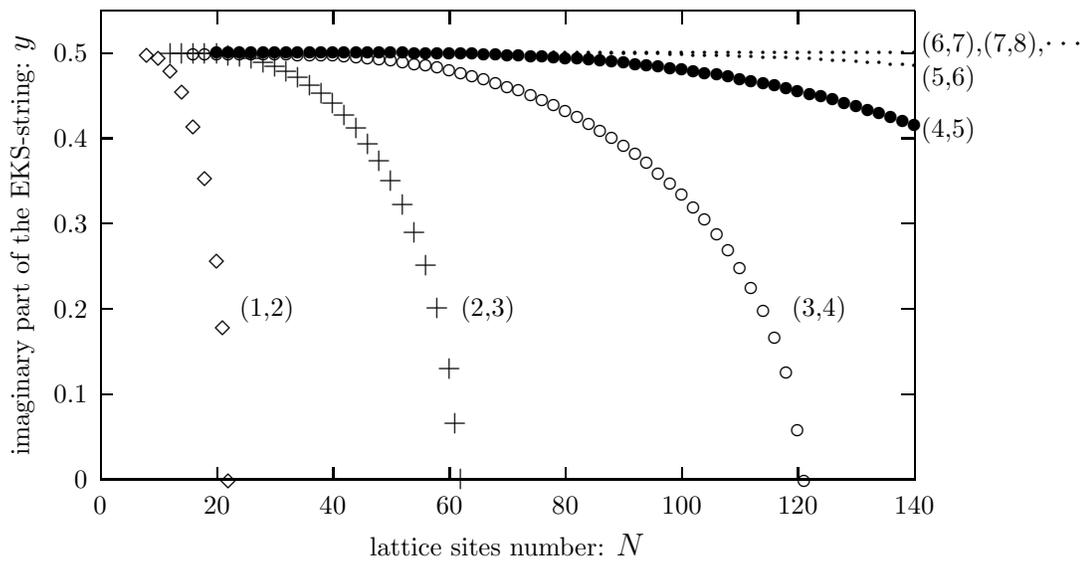
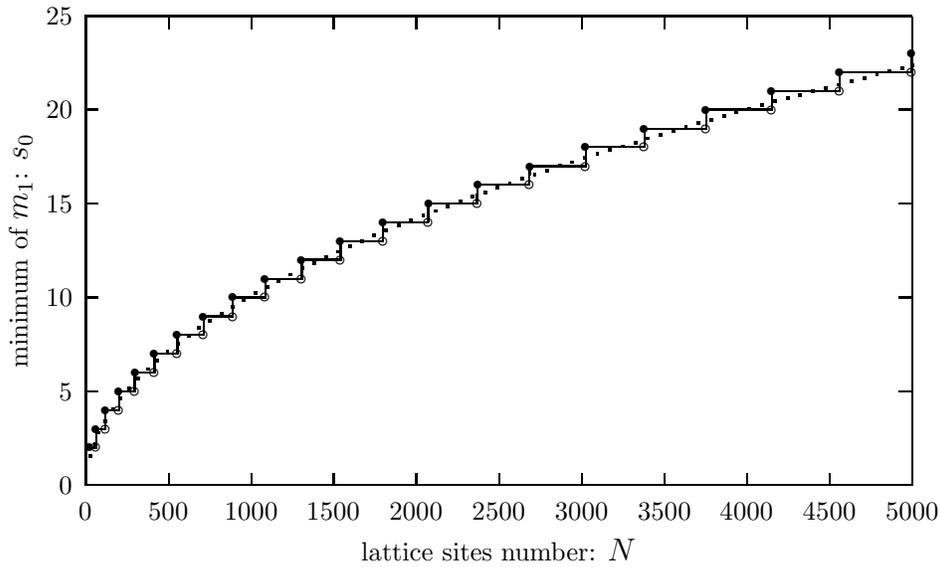
\begin{table}
\begin{center}
\caption{
The configurations and the number of the states $N_{\mbox{string}}$ of the complex solutions
for the XXX model ($\Delta=1$).
Here, we list for $m_1,m_2<N/2$ cases. The other cases are given by taking $(N-m_2,N-m_1)$.
The boldface show the missing point of the string.
}
\label{tbl_conf_string}

\vspace{2mm}
\begin{tabular}{cllc}
\hline
  $N$ & \multicolumn{2}{c}{$(m_1,m_2)$} & $N_{\mbox{string}}/2$ \rule[-3mm]{0pt}{8mm}\\\hline \hline
   8 & (1,1)  & (1,2) & 2 \rule[0pt]{0pt}{5mm}\\
  10 & (1,1),(2,2) & (1,2) & 3 \\
  \multicolumn{4}{c}{\vdots} \\
  20 & (1,1) $\sim$ (4,4) & (1,2) $\sim$ (4,5) & 8\\
  $\bm{22}$ & (1,1) $\sim$ (5,5) & (2,3) $\sim$ (4,5) & 8 \\
  \multicolumn{4}{c}{\vdots} \\
  60 & (1,1) $\sim$ (14,14) & (2,3) $\sim$ (14,15) & 27 \\
  $\bm{62}$ & (1,1) $\sim$ (15,15) & (3,4) $\sim$ (14,15) & 27 \\
  \multicolumn{4}{c}{\vdots} \\
  120 & (1,1) $\sim$ (29,29) & (3,4) $\sim$ (29,30) & 56 \\
  $\bm{121}$ & (1,1) $\sim$ (29,29) & (4,5) $\sim$ (29,30) & 55 \\
  \multicolumn{4}{c}{\vdots} \\
  199 & (1,1) $\sim$ (49,49) & (4,5) $\sim$ (48,49) & 94 \\
  $\bm{200}$ & (1,1) $\sim$ (49,49) & (5,6) $\sim$ (48,49) & 94 \\
  \multicolumn{4}{c}{\vdots} \\
  298 & (1,1) $\sim$ (74,74) & (5,6) $\sim$ (73,74) & 143 \\
  $\bm{299}$ & (1,1) $\sim$ (73,73) & (6,7) $\sim$ (73,74) & 142 \\
  \multicolumn{4}{c}{\vdots} \\
  416 & (1,1) $\sim$ (103,103) & (6,7) $\sim$ (103,104) & 201 \\
  $\bm{417}$ & (1,1) $\sim$ (103,103) & (7,8) $\sim$ (103,104) & 200 \\
  \multicolumn{4}{c}{\vdots} \rule[-4mm]{0pt}{5mm}\\
\hline
\end{tabular}

\end{center}
\end{table}

\begin{figure}
\begin{center}
\caption{
The large $N$ behavior of the V-string solutions for the XXX model ($\Delta=1$). Plot of the real part
of the string against the sites number $N$ for $(m_1,m_2)=(1,1)$ (plotted with the symbol $\circ$) and
 $(m_1,m_2)=(2,2)$ (plotted with the symbol $\bullet$).
}
\label{fig_re_string_m1}
\setlength{\unitlength}{0.240900pt}
\ifx\plotpoint\undefined\newsavebox{\plotpoint}\fi
%
\end{center}
\end{figure}

\begin{figure}
\begin{center}
\caption{
The large $N$ behavior of the EKS-string solutions for the XXX model ($\Delta=1$). Plot of the real part
of the string against the sites number $N$ for $m_2=m_1+1$ case.
}
\label{fig_re_string_m11}
\begin{minipage}{12.5cm}
\setlength{\unitlength}{0.240900pt}
\ifx\plotpoint\undefined\newsavebox{\plotpoint}\fi
\sbox{\plotpoint}{\rule[-0.200pt]{0.400pt}{0.400pt}}%
%
\end{minipage}
\end{center}
\end{figure}

\newpage
\section{Conclusion}
We have presented the systematic calculations of the large $N$ behavior in the spin
1/2 Heisenberg XXX model. Here, we have developed a new iteration method
which enables to evaluate all of the complex solutions. 
Though we have presented the calculation up to the site number $N \simeq 5000$, 
it is possible to 
extend the calculation to the larger number of $N=5000$. 
But we believe that the essential behavior of the string solutions is clarified.

In particular, we have studied the properties of the string 
solutions which are classified into two types, EKS-string and V-string.
For the EKS-string, we find a clear evidence of the violation 
of the complex solutions, and at some number $N$, they become real solutions.
We confirm the results up to $N=417$.
Initially, the EKS-string is a usual string solution, that is, the imaginary part of the
EKS-string is 1/2. However, it deviates from 1/2 when the sites number $N$ increases, 
and the imaginary part of the EKS-string becomes zero.
On the other hand, the V-string behaves in a rather normal way.
We show that the real part and the imaginary part of the V-strings behave 
like $N$ and $\sqrt{N}$, respectively, 
and the calculated results are consistent with the prediction by the $1/N$ expansion method.
Finally, the complex solutions of the XXX model in the Bethe Ansatz method are classified
 by three types of solutions for given $N$.
The first type is given by the string solution based on the string hypothesis.
 The second is given by the EKS-string whose imaginary part is lower than 1/2. The third is V-string
which behaves as $N^\alpha$ with $\alpha =1$ or $1/2$.

The EKS-string and V-string are out of the solutions based on the string hypothesis but {\it not}
 out of the Bethe Ansatz solutions.
Therefore, the number of the state which is given by the Bethe Ansatz method is unchanged. 
Further, the number of the complex solutions is less than the prediction of the string hypothesis.
But, as shown in eq.(\ref{ratio_pre}) and eq.(\ref{ratio_sthyp}), the difference is order $\sqrt{N}$.
Therefore, the thermodynamic quantity like as the entropy cannot be affected by this phenomena.

The string hypothesis has been applied to many exactly solvable models 
 and has given  physically plausible results.
However, even for the XXX model, some of the string solutions break down.
Therefore, we must carefully analyze the exactly soluble model in
 the Bethe Ansatz method with the string hypothesis \cite{woyn,des_low}.
In the massive Thirring model, the bound state is calculated using the string hypothesis \cite{ber_tha79b}.
But the recent investigation shows that it is not 
correct \cite{fuj_seki_yam,fuj_kake_tak,fuj_hir}.
We will investigate for the future issue how these violations of the string 
solutions affect the bound state problem
of the massive Thirring or XYZ model.

\end{document}